\documentclass{caosp306}
\usepackage{graphicx}
\usepackage{natbib}
\usepackage{ulem}

\articleNo{AXRO01}
\pubyear{2018}
\volume{49}
\volnumber{2}
\received{12 December 2017}
\accepted{23 January 2018} 

\def\BibTeX{{\rm B\kern-.05em{\sc i\kern-.025em b}\kern-.08em
             T\kern-.1667em\lower.7ex\hbox{E}\kern-.125emX}}

\begin{document}

\hauthor{V. Karas, Y. Kojima, and D. Kunneriath}

\title{Light rays and wave fronts in strong gravity}

\author{
        Vladim\'{\i}r Karas \inst{1} 
      \and
        Yasufumi Kojima \inst{2} 
      \and
        Devaky Kunneriath \inst{3}
   }

\institute{Astronomical Institute of the Czech Academy of Sciences,\\ Bo\v{c}n\'{\i} II 1401, CZ-14100 Prague, Czech Republic\\
	\email{vladimir.karas@cuni.cz}
         \and
Department of Physics, Hiroshima University, Higashi-Hiroshima,\\ Hiroshima 739-8526, Japan
	\and
National Radio Astronomy Observatory, 520 Edgemont Road,\\ Charlottesville 22903 (VA), USA
          }


\maketitle

\begin{abstract}
Accretion onto black holes often proceeds via an accretion disk or a temporary disk-like pattern. Variability features observed in light curves as well as theoretical models of accretion flows suggest that accretion disks tend to be inhomogeneous -- variety of substructures (clumps) emerge within the flow. Rapid orbital motion of individual clumps then modulates the observed signal in X-rays. Furthermore, changes of spectral lines and polarization properties of the observed signal (or the absence of changes) constrain the models and reveal information about general relativity (GR) effects. In this write-up we summarize the basic equations that have been employed to study light propagation near black holes and to derive the radiation signal that can be expected at a detector within the framework of geometrical optics approximation. 
\keywords{Black holes -- Accretion, accretion disks -- General relativity}
\end{abstract}

\section{Introduction}
Accreting black holes are thought to be the most likely agents driving active galactic nuclei and compact binaries.
Nonetheless, various aspects of the picture still need to be understood and several viable alternatives have been
proposed \citep[see, e.g., an overview in][]{2017FoPh...47..553E}.
Numerical and semi-analytical computations of emission continua and emission-line profiles are important
tools for verification of models with black-hole accretion disks 
\citep{1981A&A....95...18G,1989MNRAS.238..729F,1991MNRAS.250..629K,1991ApJ...376...90L,1992MNRAS.259..569K}.
Here we briefly outline the underlying ideas, believing that the disk radiation must be affected by the orbital motion of its elements.
An intrinsically narrow spectral line with a single, well-defined peak in the local rest frame co-moving with the disk material becomes broader or it can be split into more components in the frame of a distant observer. This has to be modelled in a manner consistent with the underlying continuum, taking
into account the interplay of parameters describing the system \citep{2000PASP..112.1145F}.

We adopt the standard black-hole accretion disk scenario \citep{1992apa..book.....F}, where the radiation signal emerges 
from the surface of the disk along the range of radius, $R_{\rm in}<R<R_{\rm out}$. 
Because of the temperature profile, the main contribution to the bulk emissivity of low ionization lines originates in the distant region of $\approx 10^4R_{\rm g}$,
i.e.\ near $R_{\rm out}.$\footnote{The Schwarzschild radius $R_{\rm g}$ is defined in terms of the black hole mass $M_\bullet$: $R_{\rm g}{\,\equiv\,}2G{M_\bullet}/c^2\sim3\times10^{13}\,M_8\,\rm{cm}$, where $M_8={M_\bullet}/10^8M_{\odot}$ and $M_\odot$ denotes the mass of the Sun. $R_{\rm g}$ defines the linear
size of a non-rotating black hole, whereas a maximally rotating (Kerr) black hole has the size half $R_{\rm g}$.} 
In addition, there is Doppler-boosted radiation coming from $R\approx R_{\rm in}.$ As a result, two peaks appear in the line
profile. Assuming Keplerian rotation, the orbital velocity corresponding to these peaks satisfies the relation
\begin{equation}
\frac{v_{{\rm K}\mid R=R_{\rm in}}}{v_{{\rm K}\mid R=R_{\rm out}}} \simeq \sqrt{\frac{R_{\rm out}}{R_{\rm in}}},
 \end{equation}
which is independent of the disk inclination angle. This ratio gives us the first estimate of the disk size (GR effects will modify this to certain extent). 
Typically, for $M\approx M_\odot$ one obtains
$R_{\rm out}\approx (10^4$--$10^5)R_{\rm g}$. The observer view 
angle has to be determined independently.

A direct consequence of predominantly azimuthal motion of the radiating matter is the onset of a double-peaked line core, as in a self-absorbed line. 
Near the inner rim of the accretion flow the velocity of the bulk motion grows and it becomes comparable with the speed of light. Naturally,
fast motion leads to the spectral-line broadening.
However, double-peaked profiles are only rarely observed \citep{1989ApJ...344..115C,1994ApJS...90....1E}: the lines are usually filled in. 
This is also the evidence for large $R_{\rm out}.$
On the other hand, general-relativistic light-bending and the significant frequency shifts arise at small radii, in the immediate neighbourhood of the black-hole horizon,
and they also contribute to the line asymmetry and broadening \citep{1991MNRAS.250..629K,1991ApJ...376...90L,1992MNRAS.256..679K,1995ApJ...440..108K}.

\begin{figure}[tbh!]
\centering
\includegraphics[width=\linewidth]{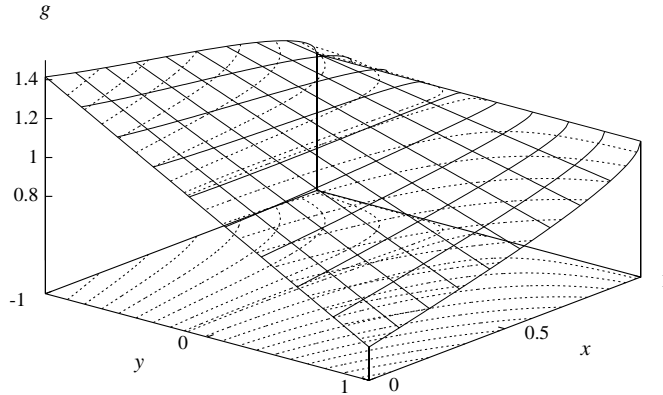}
\caption{Redshift function $g(x,y)$ in dependence on two dimensionless spatial coordinates within the disk
plane, $y=\sin\phi\sin\theta_{\rm obs}$ and $x=1-3R_{\rm g}/R.$ Different components of the overall energy shift
add up to produce energy decrease ($g>1$) or enhancement ($g<1$) in different regions of the disk. This definition of $(x,y)$ allows us 
to capture the entire disk plane from circular photon orbit at $r=3R_{\rm g}$ up to infinity, $r\rightarrow\infty$
\citep{2009ApJ...701..635M}.}
\label{g2}
\end{figure}

The Doppler effect transforms photons of radiation emitted at wavelength $\lambda_{\rm em}$ to the observed wavelength
\begin{equation}
 \lambda_{\rm obs}=\lambda_{\rm em}\,\frac{(1-\beta y)}
 {\sqrt{1-\beta^2}}\approx \lambda_{\rm em}\left(1+\frac{1}{2}\beta^2
 -\beta y\right),
\end{equation}
where $\beta=v/c,$ $y=\sin\phi\sin\theta_{\rm obs}$ $(\phi$ is the azimuthal coordinate in the disk plane and $\theta_{\rm obs}$ is the inclination
of the observer). Analogously, the gravitational redshift is
\begin{equation}
 \lambda_{\rm obs}=\frac{\lambda_{\rm em}}{1-\beta^2}.
\end{equation}
The position of the centroid of the line is independent of the disk inclination. Naturally, one cannot separate Doppler and gravitational shifts 
in a complete, self-consistent relativistic treatment of the problem. The centroid wavelength depends on inclination, when the wavelength shift, anisotropic
emissivity of the material depending on the emission angle, and gravitational focusation of light rays are properly taken into account. 

\section{Calculating the observed spectral line profiles}
The radiation originating from the gaseous elements at the inner edge of the disk experiences stronger gravitational and transverse Doppler redshift than
the radiation from regions located further out. This results in a redward asymmetry of the skewed line. In addition, the Doppler-boosted radiation
from the approaching material contributes to an enhanced peak of the blue wing of the line.  Elements of the ring emit radiation with frequency $\nu=\nu_{\rm em}$ that represents the Dirac $\delta$-function in the local rest frame. The line profile results from a superposition of individual contributions affected by 
a competition of the overall Doppler effect and the gravitational redshift. This observed profile depends on the inclination $\theta_{\rm obs}.$ For the case of a disk, the line profile can be obtained by splitting the disk into a number of concentric rings, each emitting with its own local frequency and intensity, and summing their total radiation
together.

\begin{figure}[tbh!]
\centering
\includegraphics[angle=90, width=0.7\linewidth, bb=72 150 540 620]{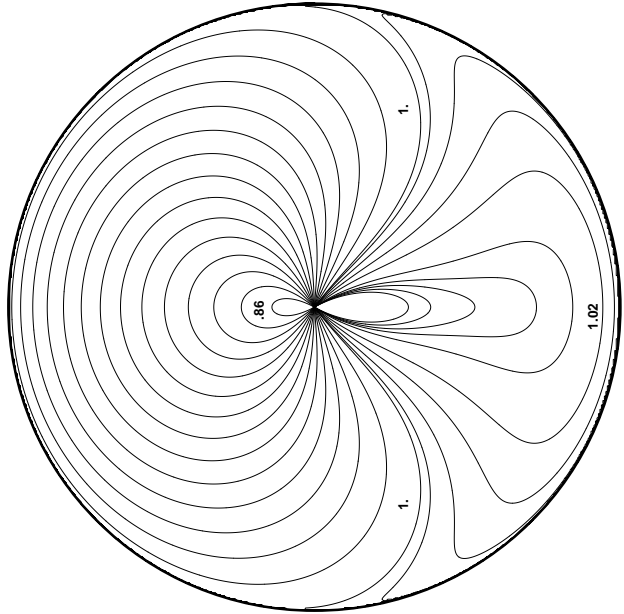}
\caption{Isocontours of $g$ in the disk surface -- a projection
along the disk axis. Radial coordinate of the circle ($x$) is defined as in the previous figure, the
azimuthal coordinate is $\phi$ here. A~distant observer is located
on the right side of the disk with a fixed value of view angle inclination, $\theta_{\rm obs}=15\deg.$
Values of $g>1$ and $g<1$ correspond to redshifted
and blueshifted radiation, respectively.}
\label{g4}
\end{figure}

Graphs of the relative frequency shift (the redshift factor)
\begin{equation}
1+z\equiv g\equiv\frac{\lambda_{\rm obs}}{\lambda_{\rm em}}
\label{defg}
\end{equation}
along the Keplerian disk surface are shown in figures \ref{g2}--\ref{g4}, taking into account general relativistic effects on the radiation.
The formula (\ref{defg}) can be rewritten in a simpler form within the pseudo-Newtonian approximation,
\begin{equation}
g=\sqrt{\frac{R-2R_{\rm g}}{\left(R-3R_{\rm g}\right)R}}
 \left(\sqrt{R-2R_{\rm g}}-y\right).
 \label{gform}
\end{equation}

\begin{figure}[tbh!]
\centering
\includegraphics[width=0.9\linewidth]{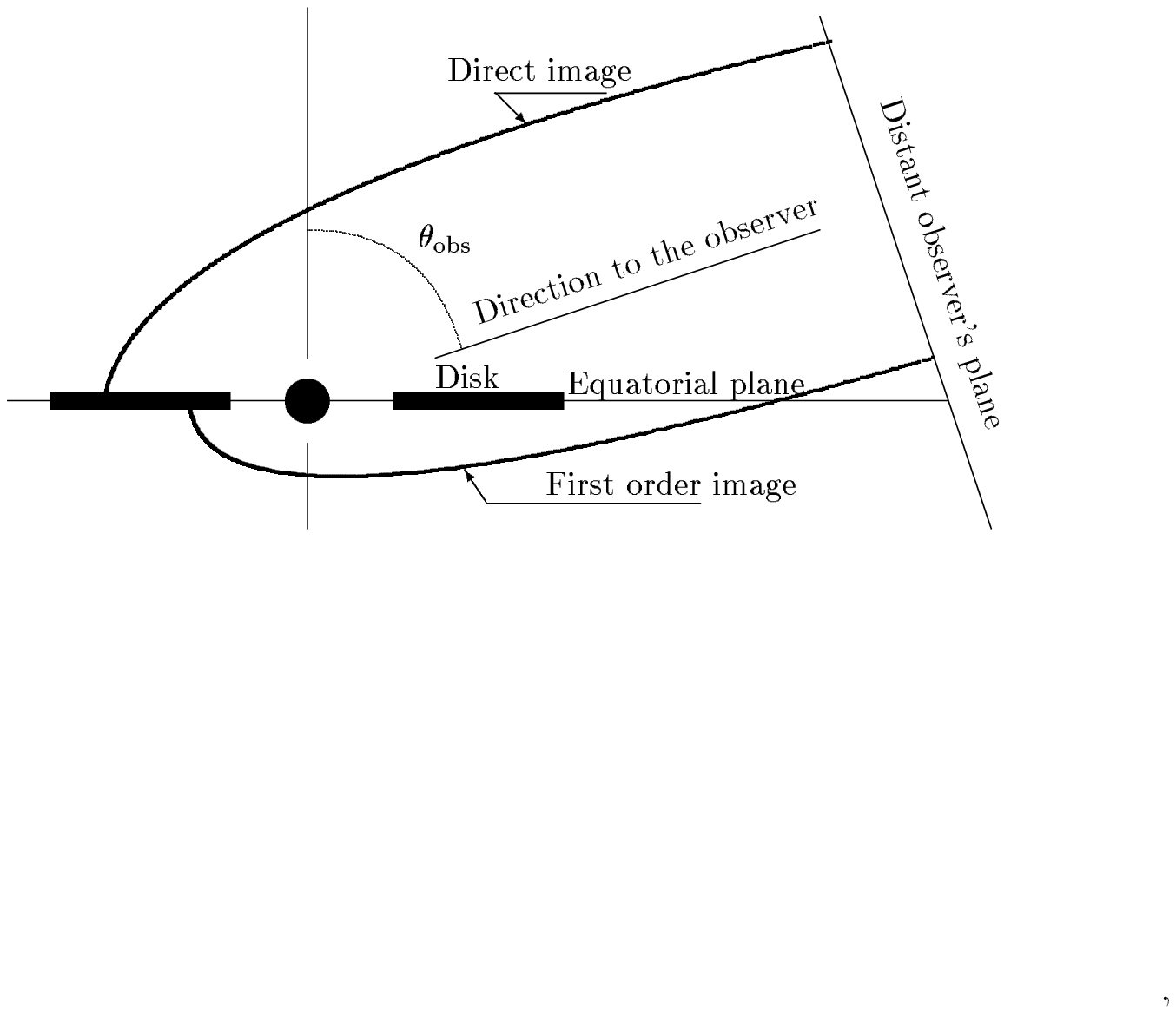}
\includegraphics[width=0.6\linewidth]{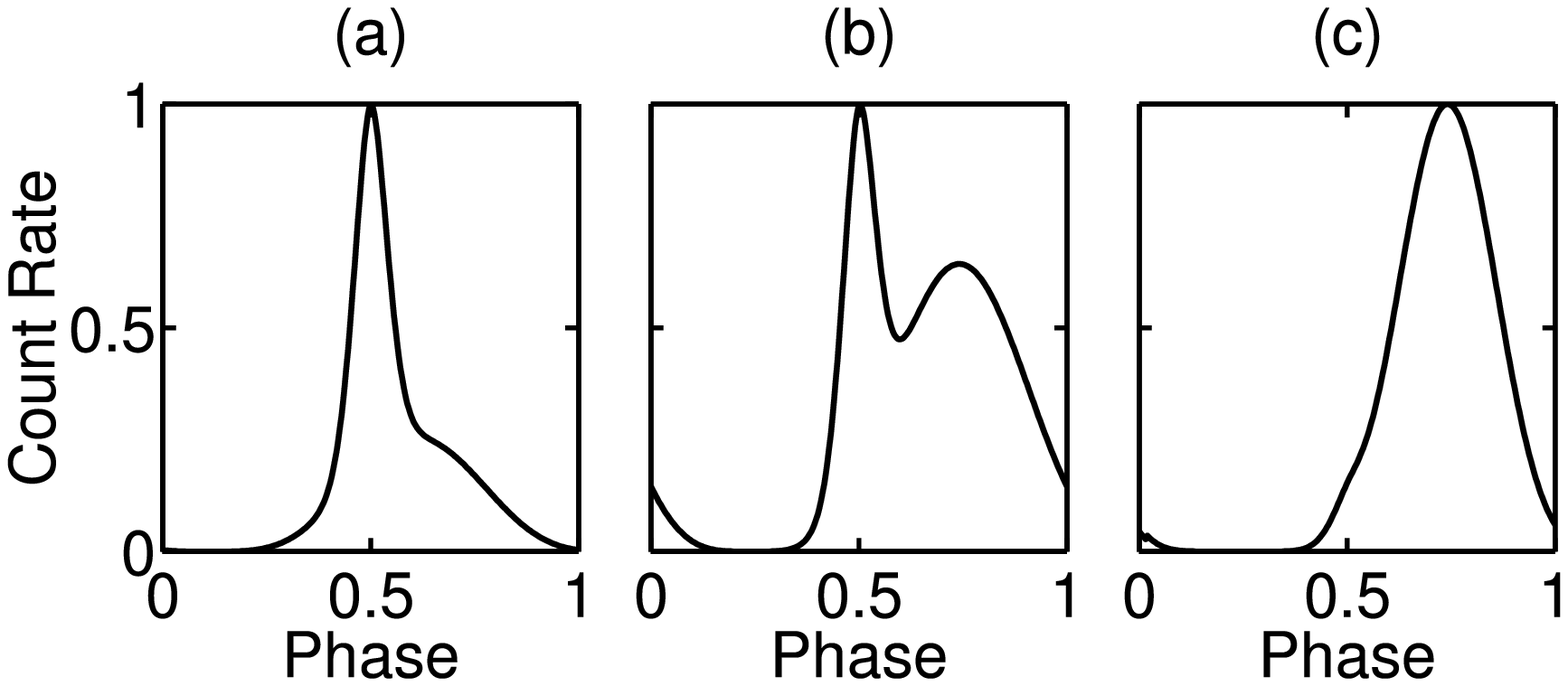}
\caption{Top: a schematic illustration of the arrangement for a simulated image of the black-hole accretion disk,
shown here in the poloidal cross-section. Bottom: Typical light-curves from a spot orbiting over one complete revolution
on the surface of an accretion disk. Parameters: (a)~$r=3R_{\rm g}$, $\theta_{\mathrm{obs}}=80$~deg
(lensing peak dominates the lightcurve near the orbital phase $0.5$);
(b)~$r=44R_{\rm g}$, $\theta_{\mathrm{obs}}=80$~deg; (c)~$r=44R_{\rm g}$,
$\theta_{\mathrm{obs}}=20$~deg (Doppler peak dominates near the phase $0.75$). For further details,
see \citet{1996PASJ...48..771K,2004ApJS..153..205D,2006AN....327..961K}.}
\label{grd1}
\end{figure}

The formula (\ref{gform}) is valid for a non-rotating black hole, however, a more complicated expression can be derived also in the rotating (Kerr) case,
where it depends on the black-hole spin (and thus it enables to constrain the black hole rotation by fitting the observed spectra).
An algorithm for the calculation of the predicted effects in spectral line profiles has been implemented in variety of modifications 
and under various assumptions
\citep[see][and references cited therein]{2006AN....327..961K}. It can be outlined in a few steps
as follows. Let us remind the reader that we assume the approximation of geometrical optics ($\lambda\ll R_{\rm g}$).

\begin{figure}[tbh!]
\centering
\includegraphics[width=0.8\linewidth]{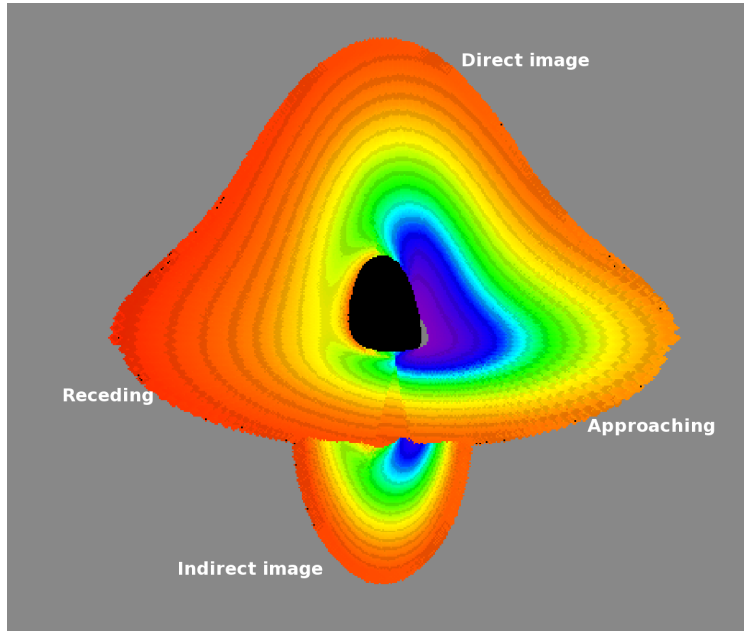}
\caption{A synthetic image of a thin disk as projected onto the detector plane and viewed by a distant observer; the case of a rotating black hole in the center (colour coded
by energy of photons at the detector plane; $a/M=0.99$, $\theta_{\rm obs}=85\;$deg). The inner rim has been set at the innermost stable circular orbit (ISCO),
while the outer radius is set at a somewhat arbitrary value.
Part of the direct image from the region of the disk behind the central object with respect to the observer acquires a significant
distortion and it appears to be elevated more above the equatorial plane. 
In addition, the observer can also see the first order (indirect) image, which has been formed by the rays crossing the equatorial
plane once (in the bottom part of the picture). The contribution of the first and the higher order images to the total 
radiation flux depends on inclination of the distant
observer, optical thickness of the disk, and a number of other parameters. The colour scale ranges from $g=0.6$ (blue, the approaching side) 
to $g=1.4$ (red, the receding part),
corresponding to energy blueshift and redshift of the emerging photons, respectively; see fig.\ \ref{fig5} for a different representation of
the redshift function across the disk plane.}
\label{grd}
\end{figure}

Firstly,  we calculate (or define) the disk surface, $z\equiv z(R),$ and determine the intensity of radiation $I^{\rm R}(R)$ which is emitted from the surface as a function of
radius, frequency of radiation and direction with respect to the disk surface in the frame co-rotating with the disk material. As we saw earlier, in the simplified
formulation of the standard model one assumes thermal radiation with an isotropic distribution in all directions arising from the equatorial plane, $z=0$, 
but the problem becomes much more complex 
if it is to be solved self-consistently with the equation of radiative transfer. In vacuum, wave fronts of emerging radiation can be solved via the eikonal equation,
which in the Schwarzschild metric reads \citep{1977PhRvD..16..933H}
\begin{equation}
-\left(1-\frac{2M}{r}\right)(\psi_{,r})^2+\left(1-\frac{2M}{r}\right)^{-1}(\psi_{,t})^2-r^{-2}(\psi_{,\phi})^2=0.
\label{eikonal}
\end{equation}
Eq.\ (\ref{eikonal}) can be solved by separation of variables, $\psi(t,r,\phi)\equiv R(r)+\alpha\phi-\omega t$. 
Photon propagation in the vacuum curved
spacetime can be formally represented in terms of wave fronts $\psi={\rm const}$ distorted by the influence of material media in the flat spacetime.
Also in Kerr metric the wave fronts can be defined and solved analytically, however, the structure of the eikonal equation is more complicated because of frame dragging.

\begin{figure}[tbh!]
\centering
\includegraphics[width=0.99\linewidth]{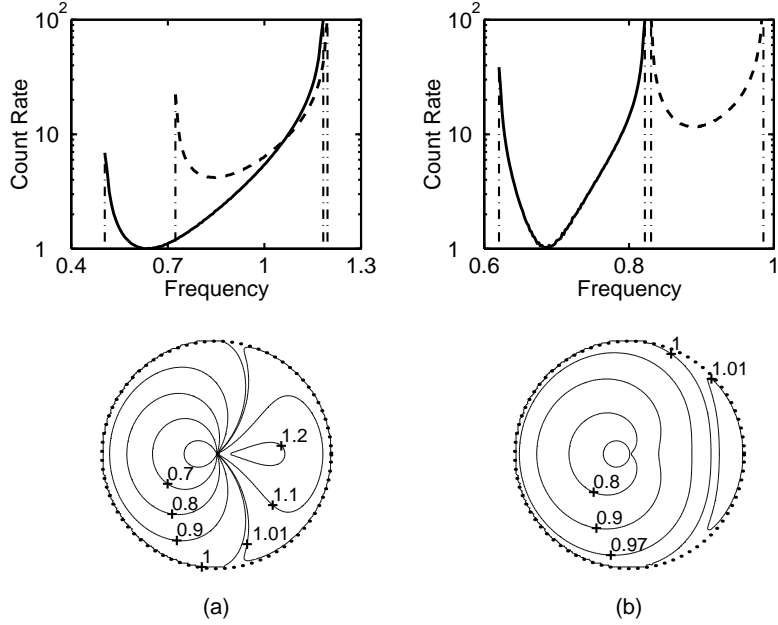}
\caption{The range and the distribution of the redshift function across the equatorial
accretion disk. Lower panels: lines of equal value of frequency shift
$g$ on a disk
in Keplerian rotation around a Schwarzschild black hole,
constructed in a way similar to fig.\ \ref{g4}:
an observer looks from the bottom of the page $(\phi=0)$.
and both plots are drawn with the radial coordinate $x\equiv1-3\,R_{\rm g}/r$ ($0\leq x\leq1$). 
Values of $g$-factor are indicated in the figure.
The whole equatorial plane outside the inner edge of the disk is thus
captured in this illustration; dotted circles correspond to $x=1$.
Observer inclination is (a) $85\,$deg (almost edge-on view), and (b) $20\,$deg
(almost pole-on view). Upper panels:
The corresponding observed photon count rates, i.e., the incoming photon flux (in arbitrary units)
from a point-like source orbiting in the disk plane at
$r_{\rm em}=3R_{\rm g}$ (solid line) and $r_{\rm em}=8R_{\rm g}$ (dashed line),
averaged over the entire revolution. Effectively, the shape represents an observed profile
of an narrow emission ring of a given radius emitting intrinsically at a single frequency
$\nu=\nu_{\rm em}$.
This shows how the photon energy is influenced by strong gravity of the black hole 
at different location of the accretion disk. 
The count rate is in arbitrary units and the frequency is normalized to the unit
emission frequency \citep[$\nu_{\rm em}=1$; see][for further details]{1996PASJ...48..771K}.}
\label{fig5}
\end{figure}

Let us note at this point that for definiteness of examples and illustration purposes we adopt the standard (relativistic) black-hole accretion disk scenario \citep{1992apa..book.....F,2008bhad.book.....K}, 
where the locally emitted spectrum is  described by multi-temperature thermal radiation (a superposition of black body spectra with the temperature changing as a function
of radius). Also, we remind the reader that our discussion concerns only the 
gravitational effects on the light rays that propagate through empty (curved) space-time of the black hole. In the other words, the entire volume above the accretion disk 
is treated as optically thin. A more realistic description will require to take the scattering effects on the photon propagation into account in the disk corona.

Secondly, we have to specify location of a distant observer with respect to the disk axis. In the usual notation, inclination angle of $\theta_{\rm obs}=90\deg$ means edge-on
view while $\theta_{\rm obs}=0\deg$ means pole-on view. Distant observer's plane (i.e., a  ``detector plane'') is perpendicular to the direction $\theta=\theta_{\rm obs}$ at an 
infinite spatial distance from the source. One has to choose a convenient 
integration grid within the disk surface if the integration is to be carried out numerically.

Each grid point (image pixel) determines unambiguously a photon ray that crosses the plane of the observer perpendicularly (fig.\ \ref{grd1}). 
We calculate intersections of 
these rays with the disk surface $z\equiv z(R)$. Photon trajectories (null geodesics) will not be straight lines in space 
if effects of general theory of relativity are taken into 
account but this fact poses only a technical complication in calculating the intersections rather than a principal difference. Therefore, thirdly, we
transform radiation intensity from the local disk frame,
$I^{\rm R}(R,z(R);\,$$\nu^{\rm R},\mu^{\rm R}),$ to the observer's laboratory frame,
$I^{\rm L}(R,z(R);\,$$\nu^{\rm L},\mu^{\rm L})$, and propagate the intensity to the observer's plane.
As mentioned above, a typical double-horn profile arises in the observed spectral line (although only one peak may be seen
for some parameter values, namely, the inclination angle and the emission radius).
Finally, we calculate the total observed flux of radiation by collecting photons,
\begin{equation}
 F^{\rm L}(\nu^{\rm L})_{\mid\theta=\theta_{\rm obs}}=
 \int_{\lower 3pt\hbox{\footnotesize (Over observer's plane)}}
 \hspace*{-17ex}I^{\rm L}(\nu^{\rm L})_{\mid R\rightarrow\infty,\theta=
 \theta_{\rm obs}}\,dS.
\end{equation}

The disk emits at temperature decreasing as a power-law
function of radius. The resulting image is created and plotted, as shown in figure \ref{grd}.
Red colour (prevailing in the left part of the image) corresponds to
a decreased energy of incoming photons with respect to energy in the local
rest-frame attached to the disk, blue colour (prevailing in the right part of the image) corresponds to an increased
energy. Light bending distorts the shape of the disk image, in particular
when the observer's inclination is large (here it is almost edge-on, $\theta_{\rm obs}=85\deg$). Notice that the image of
the inner edge of the disk is not axially symmetric due to rotation of
the black hole (dimensionless angular momentum parameter
$a=0.99$, i.e., almost maximal rotation). This figure is a typical result from computer modelling
\citep{1988PASJ...40...15F,1979A&A....75..228L,1992MNRAS.259..569K,1993A&A...272..355V}.
The main parameters
influencing the final image are the functional dependency of the intrinsic
emissivity over the disk surface, $I(r)$, the specific angular momentum $a/M$
of the black hole, and the inclination $\theta_{\mathrm{obs}}$ of the
accretion disk with respect to observer's line of sight. 

\section{Conclusions}
We discussed effects of strong gravity on radiation that propagates from an accretion disk towards a distant
observer. Imprints of these effects provide us with a possibility to study black holes residing the center of the disk.
Having in mind the applications to present-day X-ray observations, the
energy shifts, gravitational lensing and time delays are the principal
effects which originate from General Relativity and can be tested.
GR effects become dominant in the close vicinity of the hole, where they smear the resulting signal and 
are responsible for the source spectrum and its variability.

Let us note that the black-hole accretion disk model is defined by a number of parameters. There is a certain degree of degeneracy
among the parameter values, so that some parameters cannot be constrained with sufficient confidence. This results in ambiguities that 
make the fitting procedure difficult. A partial way out is often adopted by freezing some of the parameters at their likely values, however,
a more secure solution will require to gather addition observational data about the system.

Polarimetry can be employed to gain additional information.
Unlike the case of equatorial (geometrically thin) disk, the possibility of non-negligible vertical thickness introduces
an additional degree of freedom. Polarimetric properties are known to be very sensitive to the geometry of the source and we 
will need this extra piece of information to disentangle GR effects from cumbersome local physics.
More complex non-axial geometry of the emitting region needs to
be explored, such as spiral waves propagating across the accretion disk and twisted (precessing) disks.
Strong gravity near the
central black hole is very likely the main agent shaping the overall form of the
X-ray spectral features from the inner disk.

\acknowledgements
VK acknowledges the Czech Science Foundation grant No.\ 17-16287S titled ``Oscillations and Coherent Features in Accretion Disks around Compact Objects and Their Observational Signatures''. 

\bibliography{karas_caosp-2018}

\end{document}